\documentclass[aps,twocolumn,amsmath,10pt,floatfix]{revtex4}
\usepackage{epsfig}

\begin{document}
\renewcommand{\theequation}{\arabic{section}.\arabic{equation}}

%-----------------------TITLE, AUTHORS & AFFILIATIONS---------------------------
\title{From angle-action to Cartesian coordinates: A key transformation \\ for
molecular dynamics}

\author{M.~L.~Gonz\'alez--Mart\'{\i}nez}
\email[Corresponding author, ]{E-mail: mleo@instec.cu}
\affiliation{Departamento de F\'{\i}sica General, Instituto Superior de
Tecnolog\'{\i}as y Ciencias Aplicadas, Habana 6163, Cuba}
\affiliation{Institut des Sciences Mol\'eculaires, Universit\'e Bordeaux 1,
351 Cours de la Lib\'eration, 33405 Talence Cedex, France}

\author{L.~Bonnet}
\email[Corresponding author, ]{E-mail: l.bonnet@ism.u-bordeaux1.fr}
\affiliation{Institut des Sciences Mol\'eculaires, Universit\'e Bordeaux 1,
351 Cours de la Lib\'eration, 33405 Talence Cedex, France}

\author{P.~Larr\'egaray}
\affiliation{Institut des Sciences Mol\'eculaires, Universit\'e Bordeaux 1,
351 Cours de la Lib\'eration, 33405 Talence Cedex, France}

\author{J.~-C.~Rayez}
\affiliation{Institut des Sciences Mol\'eculaires, Universit\'e Bordeaux 1,
351 Cours de la Lib\'eration, 33405 Talence Cedex, France}

\author{J.~Rubayo--Soneira}
\affiliation{Departamento de F\'{\i}sica General, Instituto Superior de
Tecnolog\'{\i}as y Ciencias Aplicadas, Habana 6163, Cuba}

\date{\today}

%----------------------------------ABSTRACT-------------------------------------
\begin{abstract}
\noindent The transformation from angle-action variables to Cartesian
coordinates is a crucial step of the (semi) classical description of bimolecular
collisions and photo-fragmentations.  The basic reason is that dynamical
conditions corresponding to experiments are ideally generated in angle-action
variables whereas the classical equations of motion are ideally solved in
Cartesian coordinates by standard numerical approaches.  To our knowledge, the
previous transformation is available in the literature only for triatomic
systems.  The goal of the present work is to derive it for polyatomic ones.
\end{abstract}

\maketitle

%---------------------------------INTRODUCTION----------------------------------
\section{INTRODUCTION}
\label{sec:introduction}
Molecular reaction dynamics studies aim at understanding chemical reactions and
inelastic collisions at the atomic scale.  In other words, this field of
research draws much of the conceptual framework in which chemical reactivity, in
a broad sense, can be thought \cite{rdlevine:05}.

Quantum state-resolved integral and differential cross sections (ICSs and DCSs),
measured in supersonic molecular beam experiments, are among the most
fundamental observables of molecular reaction dynamics.  This paper deals with
their classical mechanical description in a semi-classical spirit.

Most processes considered up to now involve three or four atoms, on purpose.
This allows both measurements at an amazing level of detail and accurate
theoretical descriptions of the observables from first principles.
Additionally, planetary atmospheres and interstellar clouds are mainly made of
small species which dynamics should be understood.

Nowadays, however, much of molecular science is polarized on larger systems,
like nano-objects or molecules of biological interest, and the natural trend in
molecular reaction dynamics is also to move towards increasing complexity.  More
and more polyatomic processes are thus under scrutiny.

State-of-the-art descriptions of state-resolved ICSs and DCSs are in principle
performed within the framework of exact quantum scattering approaches (EQS)
\cite{gnyman:00,phonvault:04,vaquilanti:04,salthorpe:05,whu:06,blepetit:06,xqzhang:07,ddefazio:08}.
However, despite the impressive progress of computer performance achieved in the
last decades, these approaches can hardly be applied to larger than three or
four-atom systems as the basis sizes necessary for converging the calculations
turn prohibitive.

A popular alternative is the quasi-classical trajectory method (QCTM)
\cite{rnporter:76,dgtruhlar:79,tdsewell:97}.  This approach is intuitive,
relatively easy to implement, much less time consuming than EQS approaches and
therefore, quite appealing for studying polyatomic processes.  The price to pay
is obviously a loss in accuracy as compared to EQS approaches.  Nevertheless,
significant advances have been made in the last few years through the
replacement of the standard binning (SB) procedure by the Gaussian weighting
(GW) one
\cite{lbonnet:97,lbannares:03,lbonnet:04a,txie:05,mleo:07b,lbonnet:08,mleo:08}.
In the SB method, each trajectory has the same statistical weight.  On the other
hand, the GW procedure consists in weighting each trajectory by a Gaussian-like
coefficient such that the closer the final actions to integer values, the larger
the coefficient.  This procedure proves to be especially efficient when few
vibrational levels are available in the final products.  Though initially
proposed on the basis of rather intuitive arguments, the GW procedure can be
shown to find its roots in classical $S$ matrix theory, the former
semi-classical approach of molecular collisions pioneered by Miller and Marcus
in the early seventies \cite{whmiller:74,jrstine:74,mschild:91}.

Central quantities of chemical reaction theory are (1) the state-to-state
reaction probabilities $P_{\boldsymbol{m n}}$, where $\boldsymbol{n}$ and
$\boldsymbol{m}$ are reagent and product quantum states, (2) the densities
$dP_{\boldsymbol{m n}}/d\theta$, where $\theta$ is the scattering angle or any
given angle of the problem and (3) the capture probabilities
$P_{\boldsymbol{n}}$ for processes involving long-lived intermediate complexes
\cite{fjaoiz:08}.  From these quantities, any state-resolved ICS and DCS can be
determined.

To calculate the previous probabilities (or density of), one must generate
classical dynamical conditions corresponding to quantum state $\boldsymbol{n}$.
Such a generation is readily performed in angle-action coordinates
\cite{whmiller:74,dmwardlaw:85,mschild:91} as these are in close correspondence
with quantum numbers.  On the other hand, angle-action variables should not be
used to run trajectories as contrary to Cartesian coordinates, they lead to
strong numerical instabilities.  The transformation from angle-action variables
to Cartesian coordinates is therefore a crucial step of QCTM.

For atom-diatom (semi) collisions, this transformation can be found in the book
by Whittaker \cite{etwhittaker:89} and in a paper by Miller \cite{whmiller:71}.
However, we have not been able to find in the literature the analogous
transformation for a generic type of collision.  The goal of the paper is to
thus to derive it.

%-----------------------------------THEORY--------------------------------------
\section{THEORY}
\label{sec:theory}
In this work, a prototype system is presented, namely a five-atom molecule made
of a triatomic (ABC) and a diatomic (DE).  The former can be used as a model for
a non-linear polyatomic fragment while the latter constitutes a simpler case
very commonly found in practice.  The transformation provided here will
therefore be relevant for a generic pair of molecular fragments, {\it e.g.}
diatom + diatom, asymmetric top + diatom, asymmetric top + asymmetric top,
etc\ldots after straightforward generalizations.  These, along with the
transformations in \cite{whmiller:71,etwhittaker:89} allow thus to treat any
case of interest.

We suppose the fragments are to be studied in the low energy regime where only
the lowest vibrational states can be populated, thus the harmonic description of
their vibrations is a reasonably accurate approximation.  Anharmonic corrections
can be introduced when necessary.

Throughout this work, the usual convention of boldfacing vector magnitudes is
used.  Cartesian frames centered on a generic point P are represented as
$(\mathrm{P},x,y,z)$.  A given vector $\boldsymbol{a}$ in such a frame will be
rewritten as $\boldsymbol{a}'$ if we refer it to $(\mathrm{P},x',y',z')$
instead.  Calligraphic letters are used for representing matrices and
second-rank tensors. Some standard transformations, {\it e.g.} that of normal
modes to Cartesian coordinates, are included for completeness.  Finally, the two
fragments, ABC and DE, are numbered 1 and 2 and so are their associated
magnitudes.

\subsection{Cartesian coordinates}
\label{sec2:Cartesian}
\begin{figure}[t]
\includegraphics[width=85mm]{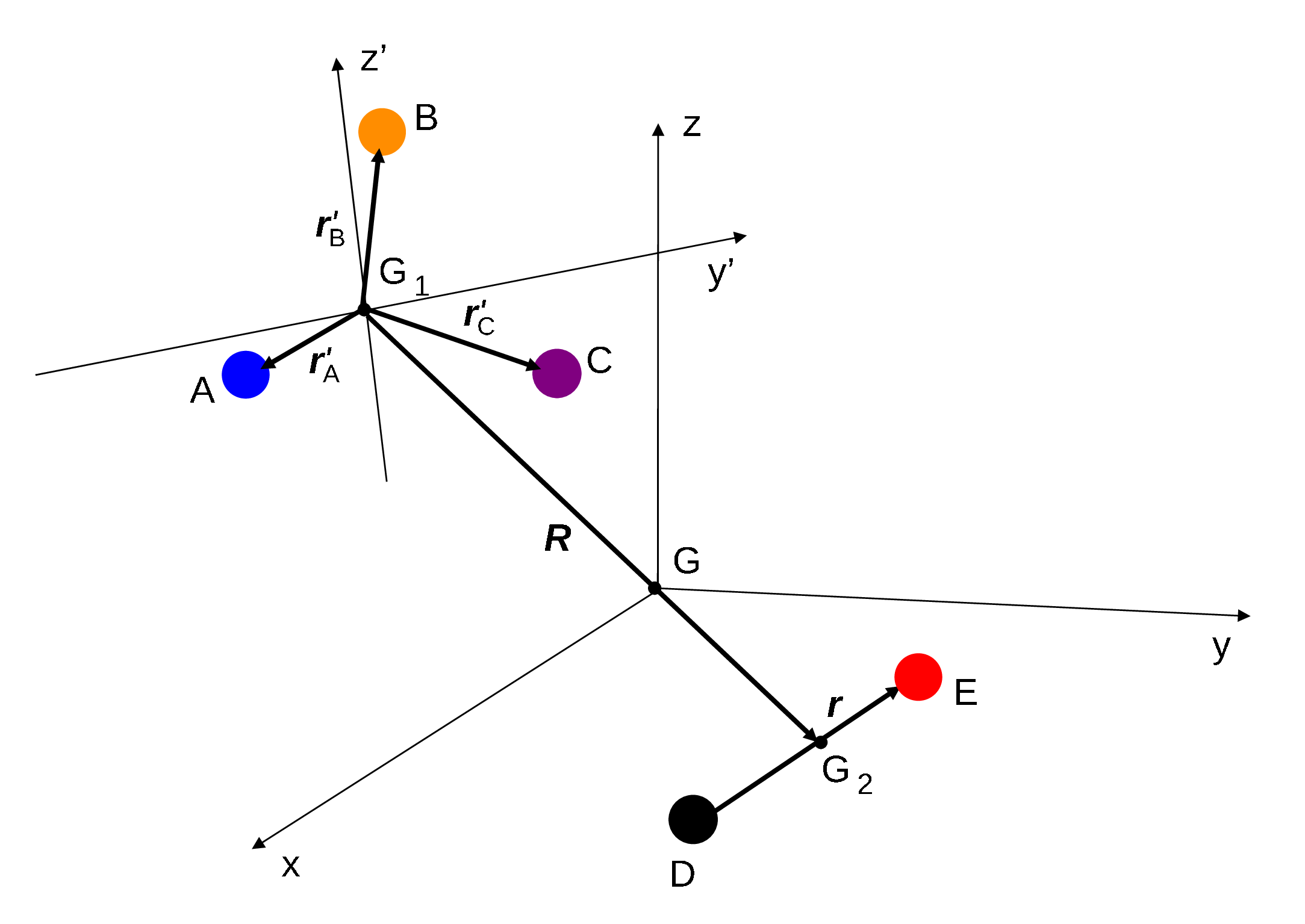}
\caption{A prototypical five-atom system represented in its center-of-mass
reference frame.  A triatomic and diatomic potential fragments are outlined.}
\label{diag:5atom}
\end{figure}

The system is schematically represented in Fig.~\ref{diag:5atom}. Three
Cartesian frames of reference are used: (1) the laboratory frame which origin is
at the molecular center of mass G and is in uniform translation so that the
total center-of-mass movement can be effectively removed, and (2, 3) the two
body-fixed, non-inertial reference frames with origins at each fragment's center
of mass, denoted G$_1$ and G$_2$.  The Cartesian coordinates to which
transformation from angle-actions is made are defined as the complete set of
nuclei positions $\boldsymbol{R}_\mathrm{X}$, in the $(\mathrm{G},x,y,z)$ space,
plus their conjugate momenta $\boldsymbol{P}_\mathrm{X}$, with
X $\in$ \{A, B,\ldots, E\}.  The total number of such coordinates yields, of
course, $6\times5=30$.

\subsection{Angle-action coordinates}
\label{sec2:angleaction}
Of the 30 variables chosen, 8 are not angle-actions, {\it i.e.} (1, 2) the
distance $R$ between the fragments centers of mass G$_1$ and G$_2$ and its
conjugate momentum $P$ \footnote{While the same letters are used as for the
general Cartesian coordinates, the subscripts in the latter will avoid any
confusion.}; and (3--8) the position and momentum vectors for the molecular
center of mass, G.  The 22 angle-action variables are thus:
\begin{list}{}{\leftmargin 1.3cm \labelwidth 1.2cm \labelsep 0.3cm}
 \item[$q_i$:] the vibrational phase of the $i$th normal mode of ABC,
$i=\overline{1,3}$.
 \item[$\hbar x_i$:] the vibrational action of the $i$th normal mode of ABC,
$i=\overline{1,3}$.
 \item[$q_4$:] the vibrational phase of DE.
 \item[$\hbar x_4$:] the vibrational action of DE.
 \item[$J$:] the modulus of the total angular momentum $\boldsymbol{J}$.
 \item[$\alpha$:] the angle conjugate to $\boldsymbol{J}$.
 \item[$J_z$:] the algebraic value of the projection of $\boldsymbol{J}$ on the
laboratory $z$ axis.
 \item[$\beta$:] the angle conjugate to $J_z$.
 \item[$l$:] the modulus of the orbital angular momentum $\boldsymbol{l}$.
 \item[$\alpha_l$:] the angle conjugate to $\boldsymbol{l}$.
 \item[$j_1$:] the modulus of the rotational angular momentum $\boldsymbol{j}_1$
of ABC.
 \item[$\alpha_1$:] the angle conjugate to $\boldsymbol{j}_1$.
 \item[$j_2$:] the modulus of the rotational angular momentum $\boldsymbol{j}_2$
of DE.
 \item[$\alpha_2$:] the angle conjugate to $\boldsymbol{j}_2$.
 \item[$k$:] the modulus of the total rotational angular momentum
$\boldsymbol{k}=\boldsymbol{j}_1+\boldsymbol{j}_2$.
 \item[$\alpha_k$:] the angle conjugate to $\boldsymbol{k}$.
 \item[$\kappa_1$:] the algebraic value of the projection of $\boldsymbol{j}_1$
on one of the three axes of inertia of ABC.
 \item[$\gamma_1$:] the angle conjugate to $\boldsymbol{\kappa}_1$.
\end{list}

The six triatomic normal mode coordinates fully specify the three position
vectors $\boldsymbol{r}'_\mathrm{A}=(y'_\mathrm{A},z'_\mathrm{A})$,
$\boldsymbol{r}'_\mathrm{B}=(y'_\mathrm{B},z'_\mathrm{B})$ and
$\boldsymbol{r}'_\mathrm{C}=(y'_\mathrm{C},z'_\mathrm{C})$,
in the $(y',z')$ plane of the body-fixed $(\mathrm{G}_1,x',y',z')$ frame of ABC.
$z'$ is arbitrarily made to coincide with one of the ABC axes of inertia when it
happens to be in its equilibrium geometry.  These six normal mode coordinates
also define the three momentum vectors
$\boldsymbol{p}'_\mathrm{A}=(p_\mathrm{Ay'},p_\mathrm{Az'})$,
$\boldsymbol{p}'_\mathrm{B}=(p_\mathrm{By'},p_\mathrm{Bz'})$
and
$\boldsymbol{p}'_\mathrm{C}=(p_\mathrm{Cy'},p_\mathrm{Cz'})$,
conjugate to the three previous position vectors, {\it i.e.} 12 coordinates as
a whole.  Note that these twelve Cartesian coordinates are deduced from the six
normal modes plus six constraints due to the fact that ABC is neither in
translation nor in rotation in the $(\mathrm{G}_1,y',z')$ plane.

\begin{figure}[h]
\includegraphics[width=85mm]{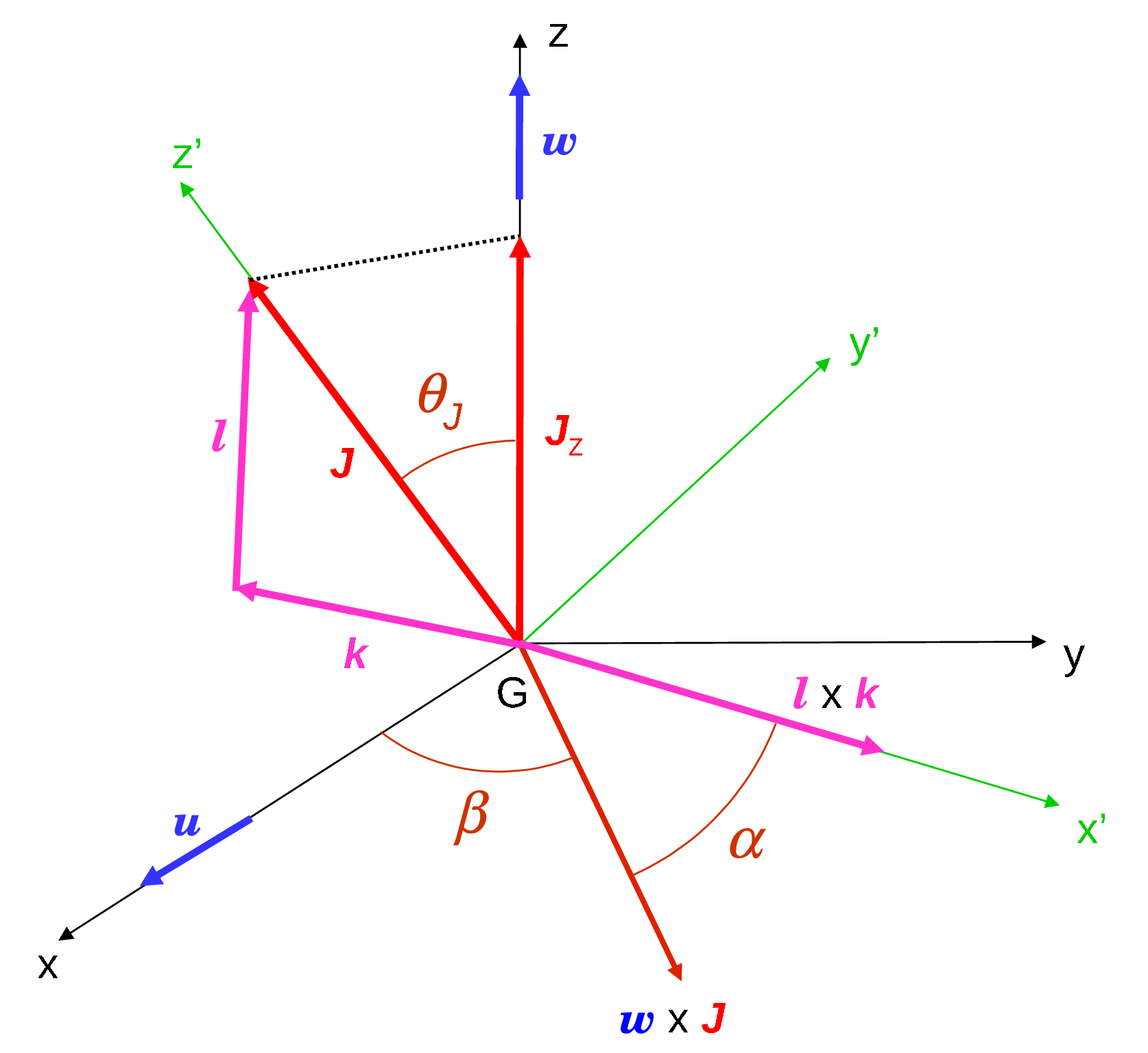}
\caption{Some angular momentum vectors and angles.}
\label{diag:Jlkgen}
\end{figure}
The total angular momentum $\boldsymbol{J}$, its $z$-component
$\boldsymbol{J}_z$, their conjugate angles $\alpha$ and $\beta$ as well as the
orbital $\boldsymbol{l}$ and total rotational $\boldsymbol{k}$ angular momenta
are represented in Fig.~\ref{diag:Jlkgen}.  The unit vectors along the $x$ and
$z$ axes are respectively denoted $\boldsymbol{u}$ and $\boldsymbol{w}$.  We
wish to emphasize here that the three axes $x'$, $y'$ and $z'$ used at this
point have nothing to do with the primed axes introduced in the previous
paragraph.  Several primed frames will be defined in the following which will be
different from each other.  $\beta$ is the angle between $\boldsymbol{u}$ and
$\boldsymbol{w}\times\boldsymbol{J}$ while $\alpha$ is the angle between
$\boldsymbol{w}\times\boldsymbol{J}$ and $\boldsymbol{l}\times\boldsymbol{k}$.

\begin{figure}[t]
\includegraphics[width=85mm]{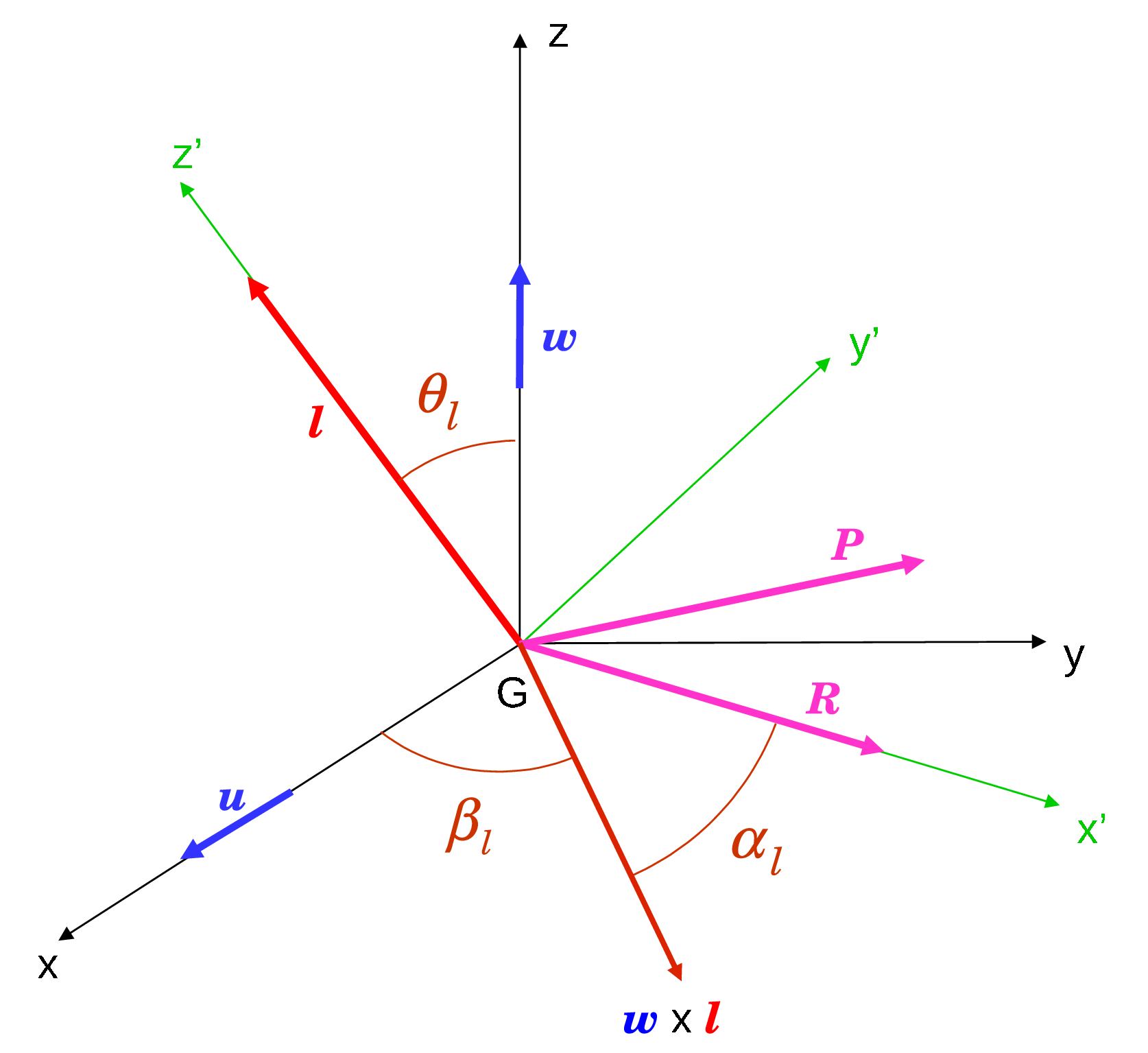}
\caption{Spatial relation among the total orbital angular momentum
$\boldsymbol{l}$, the intermolecular Jacobi vector $\boldsymbol{R}$ and its
conjugate momentum $\boldsymbol{P}$.}
\label{diag:lPR}
\end{figure}

$\boldsymbol{l}$ is represented in Fig.~\ref{diag:lPR} together with the Jacobi
vector $\boldsymbol{R}$ between G$_1$ and G$_2$.  $\alpha_l$ is the angle
between $\boldsymbol{w}\times\boldsymbol{l}$ and $\boldsymbol{R}$.  The momentum
$\boldsymbol{P}$ conjugate to $\boldsymbol{R}$ is also depicted. Like
$\boldsymbol{R}$, $\boldsymbol{P}$ lies in the plane orthogonal to
$\boldsymbol{l}$.
\begin{figure}[b]
\includegraphics[width=85mm]{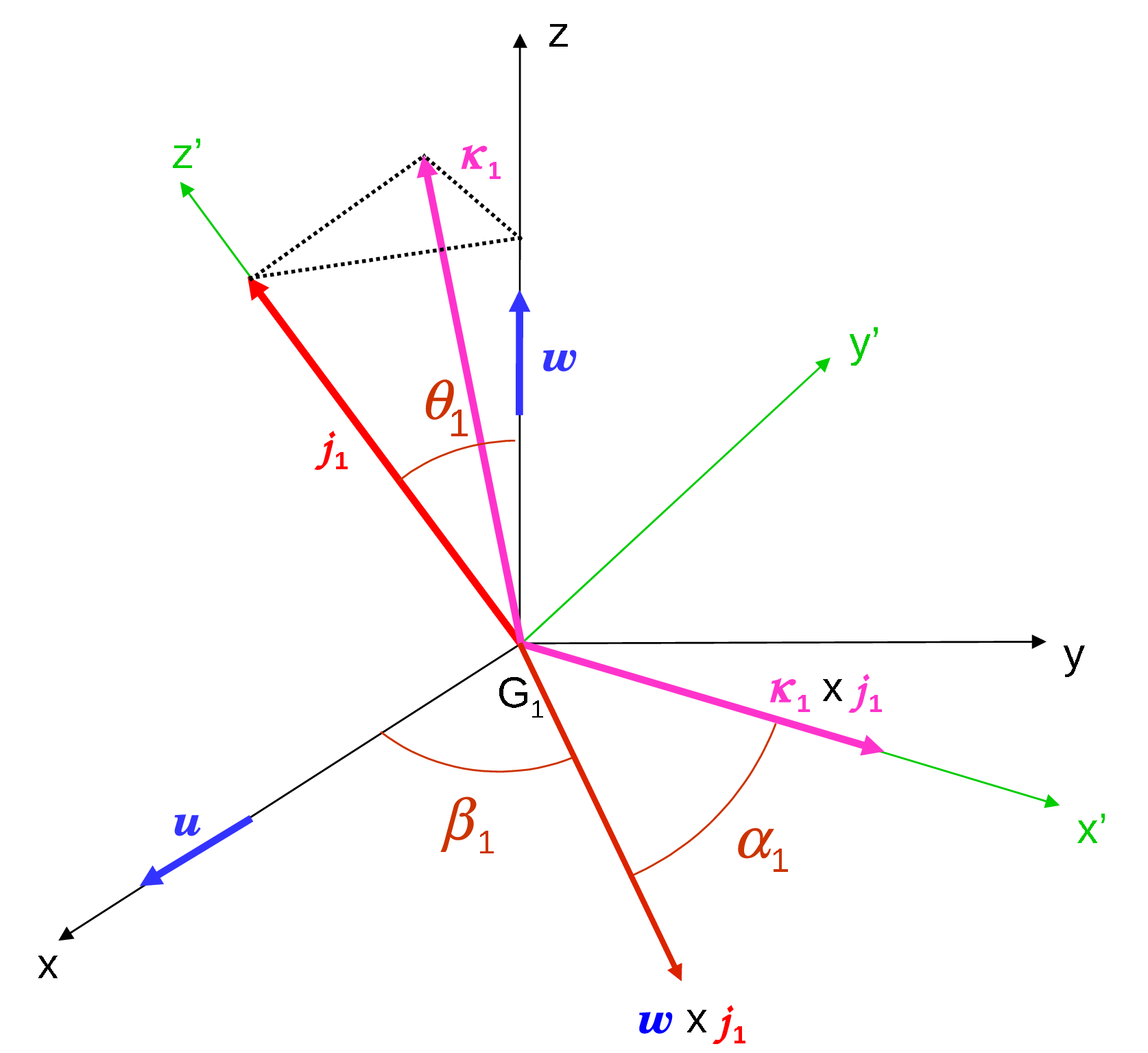}
\caption{ABC angular momentum vectors and some angles.}
\label{diag:j1k1gen}
\end{figure}

$\boldsymbol{j}_1$ is represented in Fig.~\ref{diag:j1k1gen} together with
$\boldsymbol{\kappa}_1$, defined as the projection of $\boldsymbol{j}_1$ on the
$z'$ axis of the previously specified body-fixed frame of ABC.  $\alpha_1$ is
the angle between $\boldsymbol{w}\times\boldsymbol{j}_1$ and
$\boldsymbol{\kappa}_1\times\boldsymbol{j}_1$.

\begin{figure}[t]
\includegraphics[width=85mm]{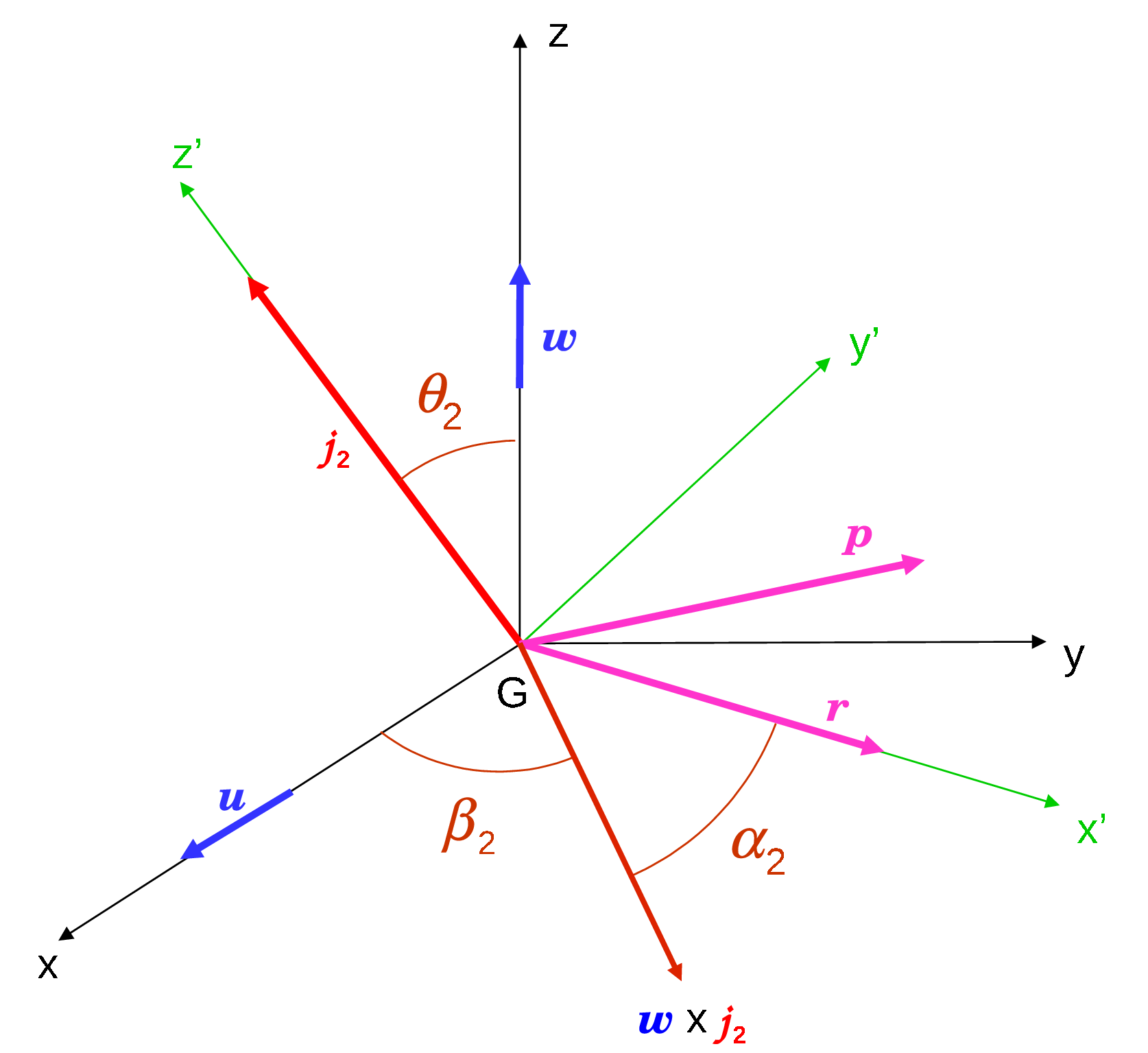}
\caption{Spatial relation among the diatomic rotational angular momentum
$\boldsymbol{j}_2$, the DE interatomic Jacobi vector $\boldsymbol{r}$ and its
conjugate momentum $\boldsymbol{p}$.}
\label{diag:j2pr}
\end{figure}

$\boldsymbol{j}_2$ is represented in Fig.~\ref{diag:j2pr} together with the
Jacobi vector $\boldsymbol{r}$ between the D and E atoms.  $\alpha_2$ is the
angle between $\boldsymbol{w}\times\boldsymbol{j}_2$ and $\boldsymbol{r}$.  The
momentum $\boldsymbol{p}$ conjugate to $\boldsymbol{r}$ is also represented.
Both $\boldsymbol{r}$ and $\boldsymbol{p}$ lie in the plane orthogonal to
$\boldsymbol{j}_2$.

\begin{figure}[b]
\includegraphics[width=85mm]{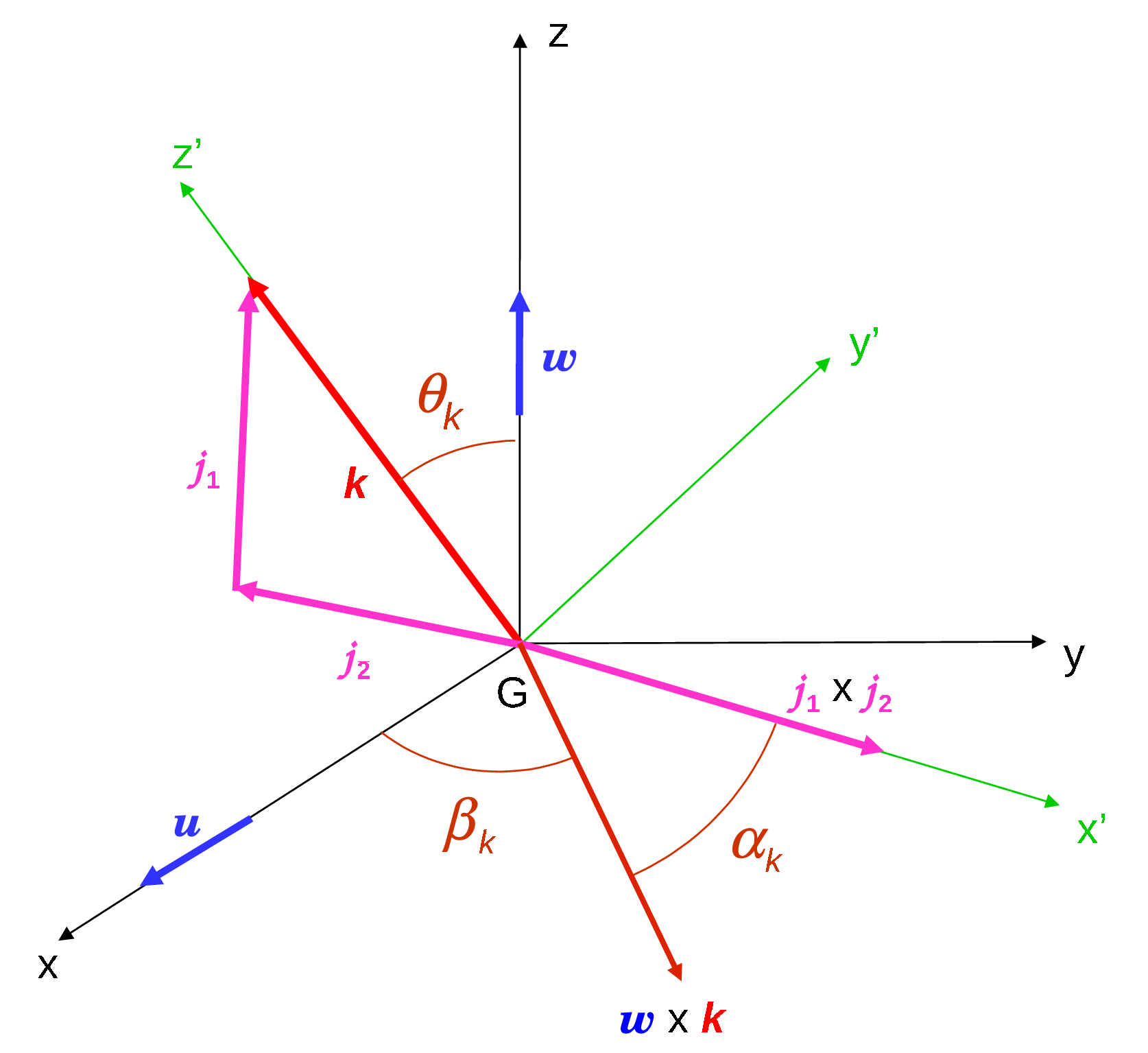}
\caption{Some angular momentum vectors and angles.}
\label{diag:kj1j2}
\end{figure}
The link between $\boldsymbol{k}$, $\boldsymbol{j}_1$ and $\boldsymbol{j}_2$ is
isomorphic to the one between $\boldsymbol{J}$, $\boldsymbol{l}$ and
$\boldsymbol{k}$, as easily seen from the comparison between
Fig.~\ref{diag:kj1j2} and Fig.~\ref{diag:Jlkgen}.  $\alpha_k$ is thus the angle
between $\boldsymbol{w}\times\boldsymbol{k}$ and
$\boldsymbol{j}_1\times\boldsymbol{j}_2$.  Calling $\boldsymbol{w}'$ the unit
vector along the $z'$ axis of the ABC body-fixed frame, the algebraic value
$\kappa_1$ equals plus (minus) $|\boldsymbol{\kappa}_1|$ when $\boldsymbol{w}'$
and $\boldsymbol{j}_1$ make an angle lower (larger) than $\pi/2$.  Finally,
$\gamma_1$ is the angle between $\boldsymbol{w}\times\boldsymbol{\kappa}_1$ and
the $x'$ axis in Fig.~\ref{diag:j1k1gen}.

\subsection{Transformation from angle-action to Cartesian coordinates}
\label{sec2:transformation}
The algorithm for computing initial conditions from the title transformation
will vary slightly according to the specific application ({\it e.g.}
unimolecular dissociation, bimolecular collision\ldots) and/or the experimental
conditions to be reflected.  The transformations, however, are intrinsically
general so we assume in what follows that all angle-action variables, as well as
$R$ and $P$, are either known or can be computed by the time they are referred
to during the process.  The transformation can be decomposed in 11 steps, each
making the subject of one of the following sections.  It is important to note
that the ordering given here is somewhat arbitrary and need for reordering may
arise in specific applications.

\subsubsection{Cartesian components of $\boldsymbol{l}$.}
\label{sec3:l}
\begin{figure}[t]
\includegraphics[width=60mm,height=65mm]{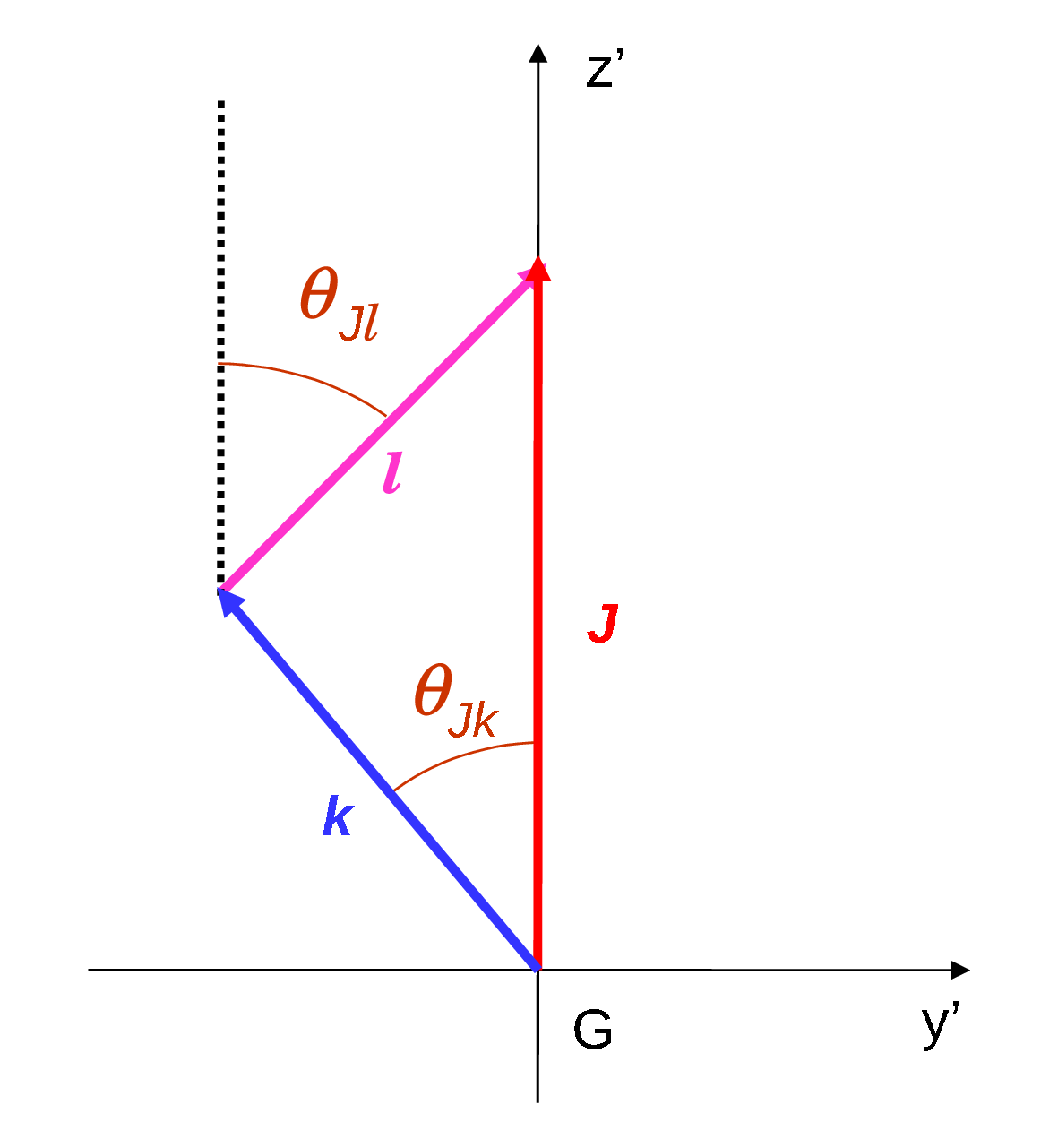}
\caption{Spatial relation between the total $\boldsymbol{J}$, orbital
$\boldsymbol{l}$ and rotational $\boldsymbol{k}$ angular momenta.}
\label{diag:Jlk}
\end{figure}

In Fig.~\ref{diag:Jlk}, the vectors $\boldsymbol{J}$, $\boldsymbol{l}$ and
$\boldsymbol{k}$ are represented in the plane $(\mathrm{G},y',z')$ as deduced
from Fig.~\ref{diag:Jlkgen}.  The relation between these angular momenta can be
written as
\begin{equation}
\boldsymbol{J}-\boldsymbol{l}=\boldsymbol{k}.
\label{eq:t1}
\end{equation}
Squaring each side of the previous equality and rearranging leads to
\begin{equation}
\cos{\theta_{Jl}}=\frac{J^2+l^2-k^2}{2Jl}.
\label{eq:t2}
\end{equation}
$l'_z$, equal to $l\cos{\theta_{Jl}}$, is thus given by
\begin{equation}
l'_z=\frac{J^2+l^2-k^2}{2J}.
\label{eq:t3}
\end{equation}
$l'_y$, equal to $l \sin{\theta_{Jl}}$, {\it i.e.}, to
$l(1-\cos^2{\theta_{Jl}})^{1/2}$ (given the convention adopted, $l'_y$ is
necessarily positive), is therefore given by
\begin{equation}
l'_y=\left[l^2-\left(\frac{J^2+l^2-k^2}{2J}\right)^2\right]^{1/2}.
\label{eq:t4}
\end{equation}
At last, $l'_x$ is zero.

$\boldsymbol{l}$ is deduced from $\boldsymbol{l}'$ by the standard Euler
rotation
\begin{equation}
\boldsymbol{l}=\mathcal{M}_3(-\beta)\mathcal{M}_1(-\theta_J)
\mathcal{M}_3(-\alpha)\,\boldsymbol{l}',
\label{eq:t5}
\end{equation}
where, for a given angle $\chi$,
\begin{equation}
\mathcal{M}_1(-\chi)=\left(\begin{array}{ccc}
                      1  &      0     &       0     \\
                      0  & \cos{\chi} & -\sin{\chi} \\
                      0  & \sin{\chi} &  \cos{\chi}
                     \end{array}\right)
\label{eq:t6}
\end{equation}
and
\begin{equation}
\mathcal{M}_3(-\chi)=\left(\begin{array}{ccc}
                      \cos{\chi} & -\sin{\chi} &  0 \\
                      \sin{\chi} &  \cos{\chi} &  0 \\
                           0     &       0     &  1
                     \end{array}\right).
\label{eq:t7}
\end{equation}
Indeed, Fig.~\ref{diag:Jlkgen} shows that one goes from $(\mathrm{G},x',y',z')$
to $(\mathrm{G},x,y,z)$ by a rotation of $-\alpha$ around the $z'$ axis followed
by a rotation of $-\theta_J$ around the resulting, `new' $x'$ axis and a final
rotation of $-\beta$ around the `new' $z'$ axis.  One may easily check that
these transformations are achieved by the $\mathcal{M}_1$ and $\mathcal{M}_3$
matrices combined as in Eq.~\ref{eq:t5}.

$\cos{\theta}_J$ is given by
\begin{equation}
\cos{\theta}_J=\frac{J_z}{J}
\label{eq:t8}
\end{equation}
and $\sin{\theta}_J$, necessarily positive as $\theta_J\in[0,\pi]$, is given by
\begin{equation}
\sin{\theta}_J=\left[1-\left(\frac{J_z}{J}\right)^2\right]^{1/2}.
\label{eq:t9}
\end{equation}

\subsubsection{Cartesian components of $\boldsymbol{R}$.}
\label{sec3:R}
\begin{figure}[b]
\includegraphics[width=85mm]{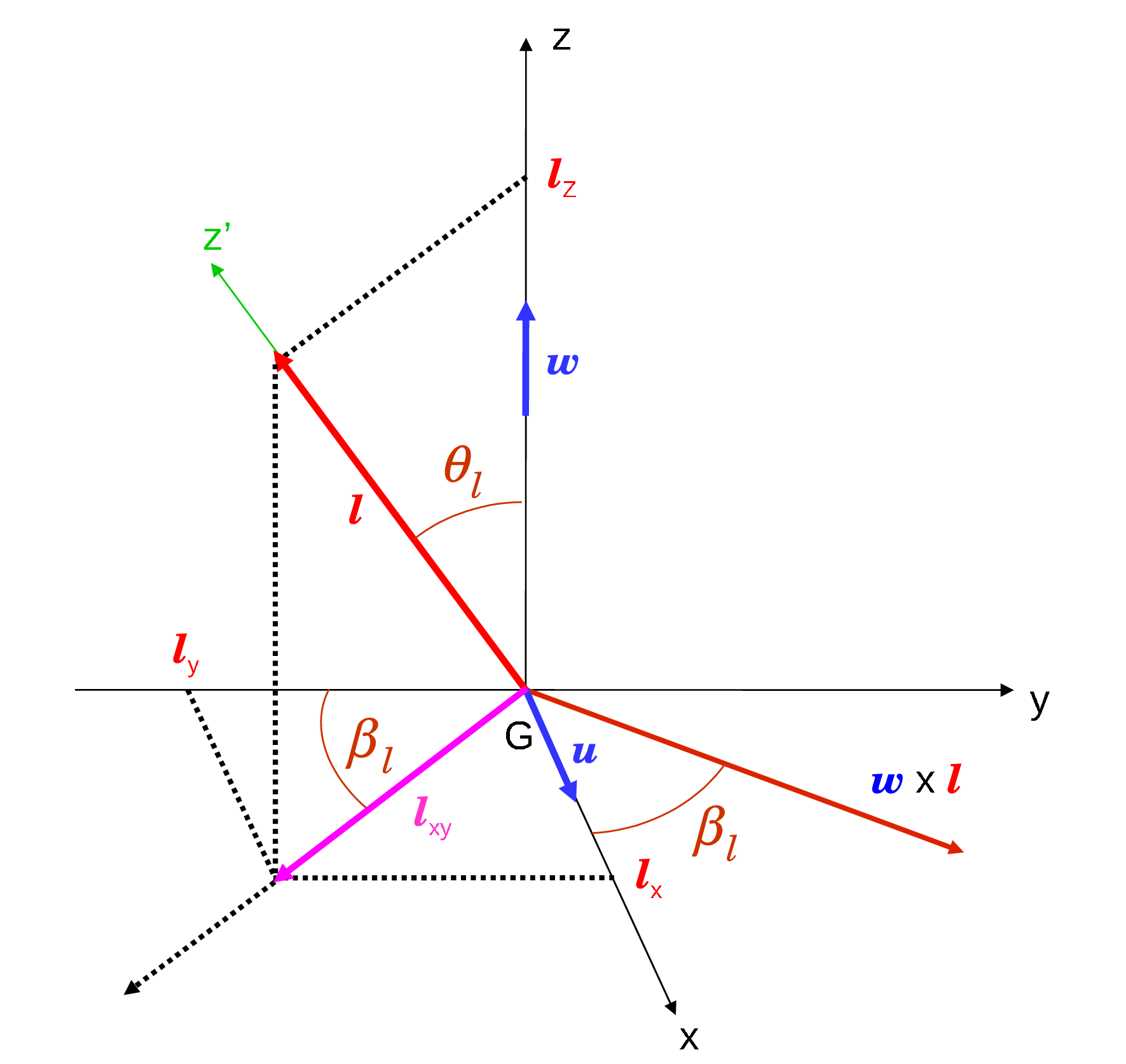}
\caption{Orbital angular momentum $\boldsymbol{l}$, its Cartesian components and
conjugate angles.}
\label{diag:lblthl}
\end{figure}
From Fig.~\ref{diag:lPR} and following the same reasoning as above,
$\boldsymbol{R}$ can be shown to satisfy
\begin{equation}
\boldsymbol{R}=\mathcal{M}_3(-\beta_l)\mathcal{M}_1(-\theta_l)
\mathcal{M}_3(-\alpha_l)\,\boldsymbol{R}',
\label{eq:t10}
\end{equation}
where $\boldsymbol{R}'$ represents the vector $(R,\,0,\,0)$.
Fig.~\ref{diag:lblthl} shows how the angles $\beta_l$ and $\theta_l$ relate to
$\boldsymbol{l}$.  $\cos{\theta}_l$ is given by
\begin{equation}
\cos{\theta}_l=\frac{l_z}{l}
\label{eq:t11}
\end{equation}
and $\sin{\theta}_l$, necessarily positive, by
\begin{equation}
\sin{\theta}_l=\left[1-\left(\frac{l_z}{l}\right)^2\right]^{1/2}.
\label{eq:t12}
\end{equation}
$\cos{\beta_l}$ is given by
\begin{equation}
\cos{\beta_l}=-\frac{l_y}{l_{xy}}
\label{eq:t13}
\end{equation}
and $\sin{\beta_l}$ by
\begin{equation}
\sin{\beta_l}=\frac{l_x}{l_{xy}},
\label{eq:t14}
\end{equation}
where
\begin{equation}
l_{xy}=(l^2-l_z^2)^{1/2}
\label{eq:t15}
\end{equation}
is the modulus of the projection of $\boldsymbol{l}$ on the $(\mathrm{G},x,y)$
plane, as depicted in Fig.~\ref{diag:lblthl}.

\subsubsection{Cartesian components of $\boldsymbol{P}$.}
\label{sec3:P}
Since $\boldsymbol{l}=\boldsymbol{R}\times\boldsymbol{P}$, $\boldsymbol{P}$ lies
in the plane $(\mathrm{G},x',y')$ of Fig.~\ref{diag:lPR}, $P'_x$ has already
been denoted $P$, $P'_y$ equals $l/R$ and $P'_z$ is zero.  $\boldsymbol{P}$ is
then obtained with
\begin{equation}
\boldsymbol{P}=\mathcal{M}_3(-\beta_l)\mathcal{M}_1(-\theta_l)
\mathcal{M}_3(-\alpha_l)\,\boldsymbol{P}'.
\label{eq:t16}
\end{equation}

\subsubsection{Cartesian components of $\boldsymbol{k}$.}
\label{sec3:k}
We still consider Fig.~\ref{diag:Jlk} and rewrite the relation between
$\boldsymbol{J}$, $\boldsymbol{l}$ and $\boldsymbol{k}$ as
\begin{equation}
\boldsymbol{J}-\boldsymbol{k}=\boldsymbol{l}.
\label{eq:t17}
\end{equation}
Squaring each side of the previous equality and rearranging leads to
\begin{equation}
\cos{\theta}_{Jk}=\frac{J^2+k^2-l^2}{2Jk}.
\label{eq:t18}
\end{equation}
$k'_z$, equal to $k\cos{\theta}_{Jk}$, is thus given by
\begin{equation}
k'_z=\frac{J^2+k^2-l^2}{2J}.
\label{eq:t19}
\end{equation}
$k'_y$, equal to $-k\sin{\theta}_{Jk}$, {\it i.e.}, to
$-k(1-cos^2\theta_{Jk})^{1/2}$ (given the convention adopted, $k'_y$ is
necessarily negative), is therefore given by
\begin{equation}
k'_y=-\left[k^2-\left(\frac{J^2+k^2-l^2}{2 J}\right)^2\right]^{1/2}
\label{eq:t20}
\end{equation}
(one may check that $k'_y$ is the just the opposite of $l'_y$).  At last, $k'_x$
is zero.  $\boldsymbol{k}$ is then obtained from $\boldsymbol{k}'$ by the same
transformation that relates $\boldsymbol{l}$ to $\boldsymbol{l}'$ (see
Eq.~\ref{eq:t5})
\begin{equation}
\boldsymbol{k}=\mathcal{M}_3(-\beta)\mathcal{M}_1(-\theta_J)
\mathcal{M}_3(-\alpha)\,\boldsymbol{k}'.
\label{eq:t21}
\end{equation}

\subsubsection{Cartesian components of $\boldsymbol{j}_1$ and
$\boldsymbol{j}_2$.}
\label{sec3:j1j2}
As already seen, the determination of $\boldsymbol{j}_1$ and $\boldsymbol{j}_2$
is in complete analogy with that of $\boldsymbol{l}$ and $\boldsymbol{k}$
(compare Fig.~\ref{diag:kj1j2} and Fig.~\ref{diag:Jlkgen}).  Following the
developments in sections \ref{sec3:l} and \ref{sec3:k}, we then arrive at
\begin{equation}
\boldsymbol{j}_i=\mathcal{M}_3(-\beta_k)\mathcal{M}_1(-\theta_k)
\mathcal{M}_3(-\alpha_k)\,\boldsymbol{j}'_i,\;\; i=1,2;
\label{eq:t22}
\end{equation}
where $j'_{1x}=j'_{2x}=0$,
\begin{equation}
j'_{1y}=\left[j_1^2-\left(\frac{k^2+j_1^2-j_2^2}{2 k}\right)^2\right]^{1/2},
\label{eq:t23}
\end{equation}
\begin{equation}
j'_{1z}=\frac{k^2+j_1^2-j_2^2}{2k},
\label{eq:t24}
\end{equation}
\begin{equation}
j'_{2y}=-j'_{1y}
\label{eq:t25}
\end{equation}
and
\begin{equation}
j'_{2z}=\frac{k^2+j_2^2-j_1^2}{2k}.
\label{eq:t26}
\end{equation}
In addition, $\cos{\theta_k}$ is given by
\begin{equation}
\cos{\theta_k}=\frac{k_z}{k}
\label{eq:t27}
\end{equation}
and $\sin{\theta_k}$ by
\begin{equation}
\sin{\theta_k}=\left[1-\left(\frac{k_z}{k}\right)^2\right]^{1/2}.
\label{eq:t28}
\end{equation}
$\cos{\beta_k}$ is given by
\begin{equation}
\cos{\beta_k}=-\frac{k_y}{k_{xy}}
\label{eq:t29}
\end{equation}
and $\sin{\beta_k}$ by
\begin{equation}
\sin{\beta_k}=\frac{k_x}{k_{xy}},
\label{eq:t30}
\end{equation}
where
\begin{equation}
k_{xy}=(k^2-k_z^2)^{1/2}
\label{eq:t31}
\end{equation}
is the modulus of the projection of $\boldsymbol{k}$ on the $(\mathrm{G},x,y)$
plane.

\subsubsection{Cartesian components of $\boldsymbol{r}$.}
\label{sec3:rp}
In the harmonic limit, the DE bond length $r$ is given in terms of $q_4$ and
$\hbar x_4$ by the expression
\begin{equation}
r=r_{\mathrm{eq}}+\left[\frac{(2x_4+1)\hbar}{\mu_2w_2}\right]^{1/2}\sin{q_4}.
\label{eq:t32}
\end{equation}
Here, $r_{\mathrm{eq}}$ is the equilibrium bond length of the diatomic, $\mu_2$
its reduced mass and $w_2$ its vibrational frequency (which is readily
determined from a quadratic fitting of its interaction potential).  Although
$x_4$ is sometimes called action, {\em stricto sensus}, this is only true in
$\hbar$ units.

The problem of the determination of $\boldsymbol{r}$ is then analogous to that
of $\boldsymbol{R}$.  From Fig.~\ref{diag:j2pr} and following section
\ref{sec3:R}, we find
\begin{equation}
\boldsymbol{r}=\mathcal{M}_3(-\beta_2)\mathcal{M}_1(-\theta_2)
\mathcal{M}_3(-\alpha_2)\,\boldsymbol{r}',
\label{eq:t33}
\end{equation}
where $\boldsymbol{r}'$ represents the vector $(r,\,0,\,0)$.  $\cos{\theta_2}$
is given by
\begin{equation}
\cos{\theta}_2=\frac{j_{2z}}{j_2}
\label{eq:t34}
\end{equation}
and $\sin{\theta_2}$ by
\begin{equation}
\sin{\theta_2}=\left[1-\left(\frac{j_{2z}}{j_2}\right)^2\right]^{1/2}.
\label{eq:t35}
\end{equation}
$\cos{\beta_2}$ is given by
\begin{equation}
\cos{\beta_2}=- \frac{j_{2y}}{j_{2xy}}
\label{eq:t36}
\end{equation}
and $\sin{\beta_2}$ by
\begin{equation}
\sin{\beta_2}=\frac{j_{2x}}{j_{2xy}},
\label{eq:t37}
\end{equation}
where
\begin{equation}
j_{2xy}=(j_2^2-j_{2z}^2)^{1/2}
\label{eq:t38}
\end{equation}
is the modulus of the projection of $\boldsymbol{j}_2$ on the $(\mathrm{G},x,y)$
plane.

\subsubsection{Cartesian components of $\boldsymbol{p}$.}
\label{sec3:prp}
Again, the problem of the determination of $\boldsymbol{p}$ is analogous to that
of the determination of $\boldsymbol{P}$.  Following section \ref{sec3:P}, we
arrive at
\begin{equation}
\boldsymbol{p}=\mathcal{M}_3(-\beta_2)\mathcal{M}_1(-\theta_2)
\mathcal{M}_3(-\alpha_2)\,\boldsymbol{p}',
\label{eq:t39}
\end{equation}
where $\boldsymbol{p}'=(p,\,j_2/r,\,0)$ and
\begin{equation}
p=\left[(2x_4+1)\hbar w_2\mu_2\right]^{1/2}\cos{q_4}
\label{eq:t40}
\end{equation}
in the harmonic approximation.

\begin{figure}[b]
\includegraphics[width=60mm,height=65mm]{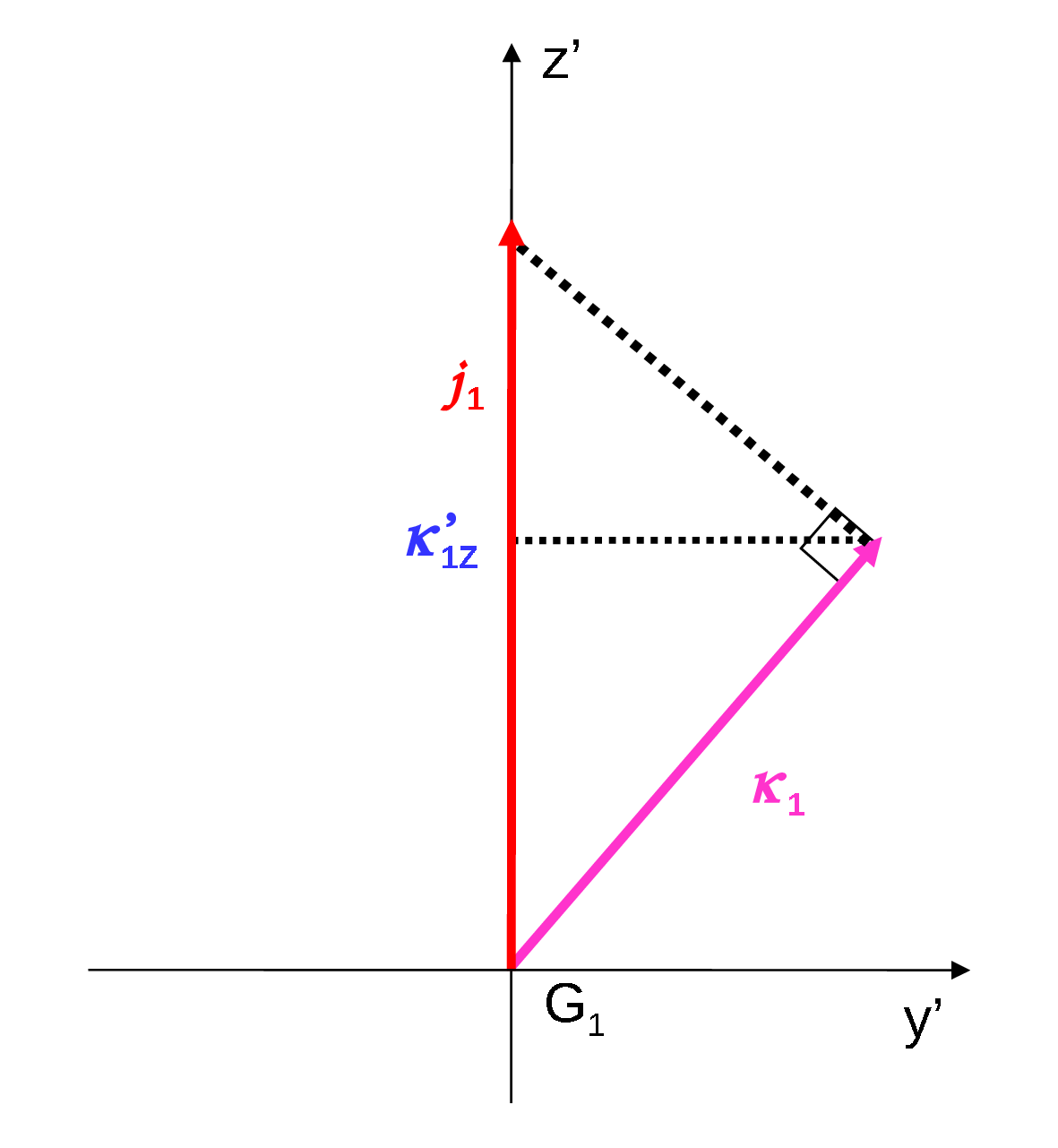}
\caption{Spatial relation between the triatomic rotational $\boldsymbol{j}_1$
and inertial-axis component $\boldsymbol{\kappa}_1$ angular momenta.}
\label{diag:j1k1}
\end{figure}
\subsubsection{Cartesian components of $\boldsymbol{\kappa}_1$.}
\label{sec3:kappa1}
$\boldsymbol{j}_1$ and $\boldsymbol{\kappa}_1$ are represented in
Fig.~\ref{diag:j1k1gen} and Fig.~\ref{diag:j1k1}.  The coordinates of
$\boldsymbol{\kappa}_1$ in $(\mathrm{G}_1,x',y',z')$ are given by
$\kappa'_{1x}=0$,
\begin{equation}
\kappa'_{1y}=|\kappa_1|\left[1-\left(\frac{\kappa_1}{j_1}\right)^2\right]^{1/2}
\label{eq:t41}
\end{equation}
and
\begin{equation}
\kappa'_{1z}=\frac{\kappa_1^2}{j_1}
\label{eq:t42}
\end{equation}
(the last equation comes from the fact that the cosine of the angle between
$\boldsymbol{\kappa}_1$ and $\boldsymbol{j}_1$ is equal both to
$\kappa'_{1z}/\kappa_1$ and $\kappa_1/j_1$, as is obvious from
Fig.~\ref{diag:j1k1}).  Proceeding as previously, we find
\begin{equation}
\boldsymbol{\kappa}_1=\mathcal{M}_3(-\beta_1)\mathcal{M}_1(-\theta_1)
\mathcal{M}_3(-\alpha_1)\,\boldsymbol{\kappa}'_1.
\label{eq:t43}
\end{equation}
$\cos{\theta_1}$ is given by
\begin{equation}
\cos{\theta_1}=\frac{j_{1z}}{j_1}
\label{eq:t44}
\end{equation}
and $\sin{\theta_1}$ by
\begin{equation}
\sin{\theta_1}=\left[1-\left(\frac{j_{1z}}{j_1}\right)^2\right]^{1/2}.
\label{eq:t45}
\end{equation}
$\cos{\beta_1}$ is given by
\begin{equation}
\cos{\beta_1}=-\frac{j_{1y}}{j_{1xy}}
\label{eq:t46}
\end{equation}
and $\sin{\beta_1}$ by
\begin{equation}
\sin{\beta_1}=\frac{j_{1x}}{j_{1xy}},
\label{eq:t47}
\end{equation}
where
\begin{equation}
j_{1xy}=(j_1^2-j_{1z}^2)^{1/2}
\label{eq:t48}
\end{equation}
is the modulus of the projection of $\boldsymbol{j}_1$ on the
$(\mathrm{G}_1,x,y)$ plane.

\subsubsection{Cartesian components of $\boldsymbol{r}_\mathrm{A}$,
$\boldsymbol{r}_\mathrm{B}$ and $\boldsymbol{r}_\mathrm{C}$.}
\label{sec3:rArBrC}
We start by determining the position vectors $\boldsymbol{r}'_\mathrm{X}$ for
\mbox{X = A, B or C} in the $(\mathrm{G}_1,x',y',z')$ frame
(Fig.~\ref{diag:5atom}).  Within the harmonic approximation, this task is
accomplished by the standard normal mode analysis \cite{tdsewell:97} (a
generalization of the procedure used in the diatomic case; compare this and the
following with sections \ref{sec3:rp} and \ref{sec3:prp}).

First, the eigenvalues $\lambda_i$ and eigenvectors $\boldsymbol{\mathcal{L}}_i$
of the Hessian matrix $\mathcal{H}$ are determined.  For an $N$-atom molecule,
six of the former correspond to the center-of-mass movement and overall rotation
and thus are theoretically zero (negligibly small in practice).  The $3N-6$
non-zero eigenvalues, associated with the molecule internal vibrational modes,
relate to their angular frequencies simply by $w_i=\lambda_i^{1/2}$.
Quasi-classical normal mode energies are then computed from the corresponding
vibrational actions as
\begin{equation}
E_i=\hbar w_i\left(x_i+\frac{1}{2}\right),
\label{eq:t49}
\end{equation}
which allows the calculation of the normal mode displacements
\begin{equation}
Q_i=\sqrt{\frac{2E_i}{\lambda_i}}\sin{q_i}.
\label{eq:t50}
\end{equation}
Cartesian mass-weighted displacements are determined with
$\boldsymbol{\eta}=\mathcal{L}\boldsymbol{Q}$, where $\mathcal{L}$ is the
eigenvector matrix and $\boldsymbol{Q}$ that of normal mode coordinates.  The
position vectors $\boldsymbol{r}'_\mathrm{X}$ are thus
\begin{equation}
\boldsymbol{r}'_\mathrm{X}=\boldsymbol{r}'_\mathrm{Xeq}+
m^{-\frac{1}{2}}_\mathrm{X}\boldsymbol{\eta}_\mathrm{X},
\label{eq:t51}
\end{equation}
where $\boldsymbol{r}'_\mathrm{Xeq}$ are the equilibrium position vectors,
$m_\mathrm{X}$ is the mass of atom X and $\boldsymbol{\eta}_\mathrm{X}$ is
extracted from $\boldsymbol{\eta}$ according to the location given to the X-atom
coordinates in $\mathcal{H}$.

\begin{figure}[t]
\includegraphics[width=85mm]{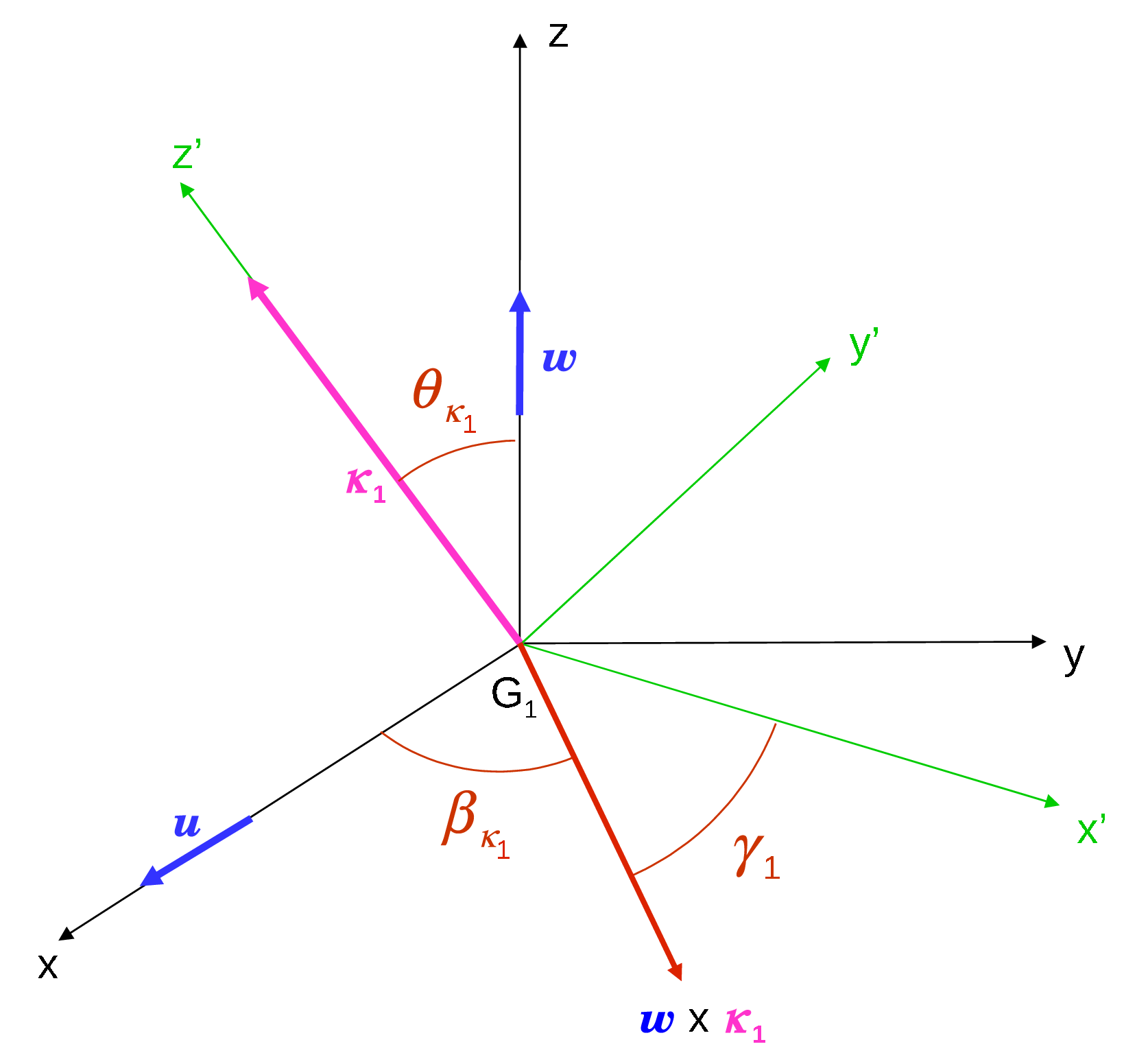}
\caption{Angular momentum vector $\boldsymbol{\kappa}_1$ and some angles.}
\label{diag:k1angles}
\end{figure}

From Fig.~\ref{diag:k1angles}, we have
\begin{equation}
\boldsymbol{r}_\mathrm{X}=\frac{\kappa_1}{|\kappa_1|}
\mathcal{M}_3(-\beta_{\kappa_1})\mathcal{M}_1(-\theta_{\kappa_1})
\mathcal{M}_3(-\gamma_1)\,\boldsymbol{r}'_\mathrm{X},
\label{eq:t52}
\end{equation}
which holds for X = A, B or C.

When $\kappa_1$ is positive, the $(\mathrm{G}_1,x',y',z')$ frames in
Fig.~\ref{diag:k1angles} and Fig.~\ref{diag:5atom} exactly coincide.  Therefore,
the dependence of $\boldsymbol{r}_\mathrm{X}$ on $\boldsymbol{r}'_\mathrm{X}$ is
of the same kind as in the previous sections.  If, on the other hand, $\kappa_1$
is negative, $(\mathrm{G}_1,x',y',z')$ in Fig.~\ref{diag:k1angles} is different
from its equivalent in Fig.~\ref{diag:5atom}.  In fact, in this case the $y'$
and $z'$ axes are oriented in the exact opposite directions as in the previous
one.  The term $\kappa_1/|\kappa_1|$, which equals $-1$, takes this difference
into account by flipping the vector $\mathcal{M}_3(-\beta_{\kappa_1})
\mathcal{M}_1(-\theta_{\kappa_1})\mathcal{M}_3(-\gamma_1)\,
\boldsymbol{r}'_\mathrm{X}$ before it is identified as
$\boldsymbol{r}_\mathrm{X}$.

In Eq.~\ref{eq:t52}, $\cos{\theta_{\kappa_1}}$ is given by
\begin{equation}
\cos{\theta_{\kappa_1}}=\frac{\kappa_{1z}}{|\kappa_1|}
\label{eq:t53}
\end{equation}
and $\sin{\theta_{\kappa_1}}$ by
\begin{equation}
\sin{\theta_{\kappa_1}}=
\left[1-\left(\frac{\kappa_{1z}}{\kappa_1}\right)^2\right]^{1/2}.
\label{eq:t54}
\end{equation}
$\cos{\beta_{\kappa_1}}$ is given by
\begin{equation}
\cos{\beta_{\kappa_1}}=-\frac{\kappa_{1y}}{\kappa_{1xy}}
\label{eq:t55}
\end{equation}
and $\sin{\beta_{\kappa_1}}$ by
\begin{equation}
\sin{\beta_{\kappa_1}}=\frac{\kappa_{1x}}{\kappa_{1xy}},
\label{eq:t56}
\end{equation}
where
\begin{equation}
\kappa_{1xy}=(\kappa_1^2-\kappa_{1z}^2)^{1/2}
\label{eq:t57}
\end{equation}
is, as usual, the modulus of the projection of $\boldsymbol{\kappa}_1$ on the
$(\mathrm{G}_1,x,y)$ plane.

\subsubsection{Cartesian components of $\boldsymbol{p}_\mathrm{A}$,
$\boldsymbol{p}_\mathrm{B}$ and $\boldsymbol{p}_\mathrm{C}$.}
\label{sec3:pApBpC}
The momenta $\boldsymbol{p}_\mathrm{X}$, with X = A, B or C, can be decomposed
into a {\em purely} translational (vibrational) and a rotational components.
Based on the very definition of the body-fixed $(\mathrm{G}_1,x,y,z)$ frame
(Fig.~\ref{diag:5atom}), the former is directly related to
$\boldsymbol{p}'_\mathrm{X}$.  To calculate these, normal-mode velocities are
first computed using the conservation of energy
\begin{equation}
\dot{Q}_i=\pm(2E_i-\lambda_iQ^2_i)^{1/2},
\label{eq:t58}
\end{equation}
the sign being selected according to the value of the vibrational phase $q_i$.
Cartesian mass-weighted velocities are thus
$\dot{\boldsymbol{\eta}}=\mathcal{L}\dot{\boldsymbol{Q}}$ from which
\begin{equation}
\boldsymbol{p}'_\mathrm{X}=m^{\frac{1}{2}}_\mathrm{X}
\dot{\boldsymbol{\eta}}_\mathrm{X}.
\label{eq:t59}
\end{equation}
It is important to stress that the anharmonicity of the real potential energy
has been deliberately neglected within the normal-mode approximation.  To
correct for its possible spurious consequences, relatively sophisticated recipes
can be used at this stage.  The reader is thus referred to the available
literature, {\it e.g.} \cite{tdsewell:97}, as it is not our objective to
reproduce them here.

The rotational component is determined in the standard fashion.  The triatomic
angular velocity is computed as
$\boldsymbol{w}_1=\mathcal{I}^{-1}\boldsymbol{j}_1$---being $\mathcal{I}$ the
inertia tensor of ABC, which can be calculated at this point since its
configuration has been determined---from which, the corresponding linear
velocities are given by
\begin{equation}
\boldsymbol{\nu}_\mathrm{X}=\boldsymbol{w}_1\times\boldsymbol{r}'_\mathrm{X}.
\label{eq:t60}
\end{equation}

Finally, the transformation relating $\boldsymbol{p}_\mathrm{X}$ and
$\boldsymbol{p}'_\mathrm{X}$ is isomorphic to Eq.~\ref{eq:t52}, so the desired
general expression for computing the former reads
\begin{eqnarray}
\boldsymbol{p}_\mathrm{X}&=&\frac{\kappa_1}{|\kappa_1|}
\mathcal{M}_3(-\beta_{\kappa_1})\mathcal{M}_1(-\theta_{\kappa_1})
\mathcal{M}_3(-\gamma_1)\,\boldsymbol{p}'_\mathrm{X}+ \nonumber \\
                         & &m_\mathrm{X}\boldsymbol{\nu}_\mathrm{X}.
\label{eq:t61}
\end{eqnarray}

\subsubsection{Nuclear positions and momenta in $(\mathrm{G},x,y,z)$.}
\label{sec3:RXPX}
At this point it is a simple task to finally express all Cartesian vectors in
the laboratory frame.  For X = A, B or C, $\boldsymbol{R}_\mathrm{X}$ is given
by the general expression
\begin{equation}
\boldsymbol{R}_\mathrm{X}=-\frac{M_2}{M_\mathrm{tot}}\,\boldsymbol{R}+
\boldsymbol{r}_\mathrm{X},
\label{eq:t62}
\end{equation}
while if X = D or E,
\begin{equation}
\boldsymbol{R}_\mathrm{X}=\frac{M_1}{M_\mathrm{tot}}\,\boldsymbol{R}+
\boldsymbol{r}_\mathrm{X}.
\label{eq:t63}
\end{equation}
In these equations, $M_i$ stands for the mass of fragment $i$ and
$M_\mathrm{tot}$ is the system total mass.

Similar relations hold for the Cartesian momenta.  For X = A, B or C, these are
computed using the general expression
\begin{equation}
\boldsymbol{P}_\mathrm{X}=-\frac{m_\mathrm{X}}{M_1}\,\boldsymbol{P}+
\boldsymbol{p}_\mathrm{X}.
\label{eq:t64}
\end{equation}
At last, the diatomic momenta are given by
\begin{equation}
\boldsymbol{P}_\mathrm{D}=\frac{m_\mathrm{D}}{M_2}\,\boldsymbol{P}-
\boldsymbol{p}
\label{eq:t65}
\end{equation}
and
\begin{equation}
\boldsymbol{P}_\mathrm{E}=\frac{m_\mathrm{E}}{M_2}\,\boldsymbol{P}+
\boldsymbol{p}.
\label{eq:t66}
\end{equation}

%----------------------------------TEST CASE------------------------------------
\section{Ketene unimolecular dissociation: A test case.}
\label{sec:testcase}
The photo-fragmentation of ketene (CH$_2$CO) has been intensively investigated
for over two decades, both experimental and theoretically  ({\it e.g.}
\cite{i-cchen:88,sjklippenstein:89,igarcia-moreno:94,sjklippenstein:96,kmforsythe:01,avkomissarov:06}).
Following photo-excitation to the $\tilde{A}^1A''$ states, the molecule
undergoes either intersystem crossing or fast internal conversion to the low
lying triplet and singlet electronic states.  From these, dissociation into
methylene and carbon monoxide occurs.  Despite the triplet threshold lies
$\sim$3150 cm$^{-1}$ below the singlet, the fact that it presents a small
barrier to dissociation---of a few cents of inverse centimeters---makes the
singlet channel statistically dominant from excess energies as low as
$\sim$100--200 cm$^{-1}$.  Such conditions make the system an effective
prototype for a {\em barrierless polyatomic} unimolecular reaction on a
{\em single} potential energy surface (PES).

In direct correspondence with the model transformation we introduced above, the
molecule constitutes a five-atom system which dissociates into a triatomic
and diatomic fragments.  Additionally, the experimental excitations are
compatible with the harmonic---normal mode---approximation for the CH$_2$ and CO
products.  In what follows we briefly report on the application of the title
transformation to the study of this process.  Full details and results will be
given in a separate work so we simply introduce it here as a corroboratory test
case.

\begin{figure}[bt]
\includegraphics[width=85mm]{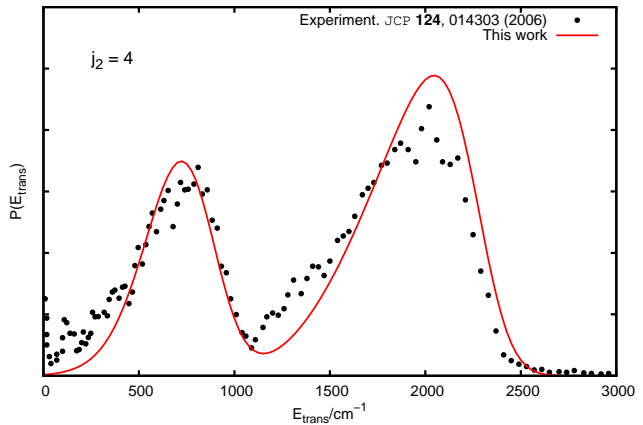}
\caption{Translational energy distribution in correlation with $j_2=4$, after
excitation with a 308 nm laser.  Comparison with the experiment.}
\label{fig:PEtJc4}
\end{figure}
In Fig.~\ref{fig:PEtJc4} we compare our calculations with the most recent
experimental results \cite{avkomissarov:06} for the products translational
energy distributions, in correlation with the rotational state of CO.  A 308 nm
laser is used in the experiment, corresponding to an excess energy of
2350 cm$^{-1}$.  The theoretical results are obtained using the so-called
{\em exit-channel corrected phase-space theory}, proposed by Hamilton and Brumer
\cite{ihamilton:85}.  This method basically consist in generating microcanonical
initial conditions {\em at the products} and then propagate the trajectories
backwards in time, the statistics being performed with those reaching the inner
transition state (TS).  The photo-excited ketene molecule is supposed to be long
lived prior to its fragmentation, thereby justifying the use of a microcanonical
distribution.  We employed the high-level {\em ab initio} PES and transition
state locations recently reported \cite{sjklippenstein:96}.

The theoretical predictions are in very good agreement with the experiment, as
can be seen in Fig.~\ref{fig:PEtJc4}.  The curve has been artificially smoothed
by using a convolution with an `apparatus' function, {\it i.e.}
\begin{equation}
P(E_\mathrm{trans})=\int{P(E')e^{-\beta(E'-E_\mathrm{trans})^2}dE'}
\end{equation}
to recover the experimental tails.  The two peaks correlating with the $v_2=0$
and $v_2=1$ CH$_2$ scissor-mode states are fairly well reproduced.

In order to further verify the validity of the transformation provided, we have
calculated the determinant of the Jacobian matrix.  The original is not a square
matrix, with dimensions 30$\times$24.  Therefore, for being able to calculate
the determinant we introduced an additional transformation to a set of Jacobi
coordinates, from which the (null) center-of-mass coordinates and momenta are
later removed.  The calculation starts with the transformation from
angle-actions to Cartesian and from these to Jacobi coordinates.  The
center-of-mass Jacobi vectors are then removed and the determinant of the
resulting 24$\times$24 Jacobian matrix, from angle-actions to (reduced) Jacobi
coordinates, is computed.  We confirmed that it yields 1 within numerical
accuracy.

%--------------------SUMMARY, CONCLUSIONS & ACKNOWLEDGMENTS---------------------
\section{SUMMARY AND CONCLUSIONS}
\label{sec:summary}
We have presented the transformation from angle-action to Cartesian coordinates,
for polyatomic systems.  In the quasi and semi-classical approaches, this
provides an expeditious way to generate initial conditions in close
correspondence with nowadays experiments and yet, solve the equations of motion
using the `ideal' Cartesian coordinates.  The methodology and expressions
provided here can either be directly used or straightforwardly generalized to
deal with any case of interest, ranging from the study of bimolecular collisions
to polyatomic unimolecular dissociations.

Preliminary results of the particular application to the study of the
unimolecular dissociation of ketene in the singlet electronic state, have been
discussed.  A very good agreement is observed between the experimental values
and theoretical predictions for correlated translational energy distributions.
The validity of the transformation have been further verified by numerical
computation of the determinant of the Jacobian matrix, which yields unity within
reasonable accuracy.

\section*{ACKNOWLEDGMENTS}
Support from an Inter-University Agreement on International Joint Doctorate
Supervision between the Instituto Superior de Tecnolog\'{\i}as y Ciencias
Aplicadas, Cuba and the Universit\'e Bordeaux 1, France, as well as the PNAP/7/3
project of the Cuban institution, are gratefully acknowledged.

%----------------------------------REFERENCES-----------------------------------

\end{document}